\documentclass[twocolumn,aps]{revtex4}

\usepackage{graphicx}

\begin{document}

\bibliographystyle{apsrev}

\title{Formation, Manipulation, and Elasticity Measurement of a Nanometric Column of Water
Molecules}

\author{H. Choe,$^{1,2}$ M.-H. Hong,$^{1,2}$ Y. Seo,$^{2}$ K. Lee,$^{1}$
G. Kim,$^{1}$ Y. Cho,$^{1}$ J. Ihm,$^{1}$ and W.
Jhe$^{1,2}$\footnote{Corresponding author: whjhe@snu.ac.kr}}

\affiliation{$^{1}$School of Physics and $^2$Center for Near-field
Atom-photon Technology, Seoul National University, Seoul 151-747,
South Korea.}

\begin{abstract}

Nanometer-sized columns of condensed water molecules are created
by an atomic-resolution force microscope operated in ambient
conditions. Unusual stepwise decrease of the force gradient
associated with the thin water bridge in the tip-substrate gap is
observed during its stretch, exhibiting regularity in step
heights ($\approx$ 0.5 N/m) and plateau lengths ($\approx$ 1 nm).
Such "quantized" elasticity is indicative of the atomic-scale
stick-slip at the tip-water interface. A
thermodynamic-instability-induced rupture of the water meniscus
(5-nm long and 2.6-nm wide) is also found. This work opens a
high-resolution study of the structure and the interface dynamics
of a nanometric aqueous column.


\end{abstract}
\date{\today}
\maketitle

Water is one of the most important substances of life and has
been studied extensively for hundreds of years. Nonetheless, it
is still quite a unique matter that keeps surprising us and
exhibits peculiarities, in particular, when confined in a
nanometric configuration. For example, water and simple organic
liquids exhibit solid-like orderedness in molecularly thin films
\cite{RLK01,HXOS95,IMH88}. Water molecules inside hydrophobic
nanotubes manifest phases of ice that are not found under bulk
conditions \cite{HRN01}. However, since bulk water possesses only
short-range order \cite{KS94} and water molecules move
incessantly, it is usually difficult to experimentally
investigate novel features of confined water structures other
than thin films.

In this Letter, we have employed an atomic-resolution noncontact
atomic force microscope (AFM) in air \cite{SCJ03} and achieved the
spontaneous formation of a nanometric liquid column consisting of
thousands of water molecules. We also have performed the
sensitive measurement of the elastic property (or the force
gradient) of the thin water column during its mechanical stretch.
We have thereby demonstrated several novel phenomena: (i) the
unusual stepwise decrease of the force gradient, associated with
the atomic-scale stick-slip on the AFM-tip surface, (ii) the
abrupt rupture of the thin water meniscus due to the thermodynamic
instability of the liquid-vapor interface, and (iii) the
manipulation of the thin aqueous column by repeated
stretch-relaxation cycles, revealing the atomic-scale contact
angle hysteresis.

Water molecules in ambient conditions produce a nanoscale water
meniscus between a hydrophilic Si tip and a hydrophilic mica
substrate, when capillary condensation occurs as the stiff AFM tip
approaches the substrate within a nanometric distance (Fig. 1).
Once the thin aqueous column is formed, it is stretched
vertically upward by subsequent retraction of the tip. As the
molecular water bridge of sub-zeptoliter (zepto = $10^{-21}$)
volume is elongated thereby, the force gradient associated with
the elasticity of the system is measured by an extremely small
amplitude-modulation operation of AFM \cite{GP02,GSBL03}.

Figure 1 presents the schematics of a home-built AFM setup used
for formation of a nanometric water column by capillary
condensation as well as for simultaneous measurement of the force
gradient of the elongated water meniscus, obtained at a given
relative humidity (RH) and at room temperature (21 $^{\rm o}$C). A
small-radius ($\approx$ 10 nm) Si tip, having a nanometric
discrete structure of atomic layers, is attached to a
high-frequency $f_0$ (= 1,002,198 Hz) quartz crystal oscillator
that has a very high stiffness $k_0$ ($\approx 5.4 \times 10^5$
N/m) and quality factor $Q$ ($\approx$ 10$^4$). These parameters,
under ambient conditions, allow the tip to be operated in
extremely small amplitude-modulation ($\approx$ 0.01 nm),
noncontact, tapping mode suitable for high-resolution
force-gradient measurement.

In contrast to a conventional micro-fabricated cantilever-based
AFM used either in contact or noncontact operation mode, our AFM
tip is stiff enough to pull the condensed water molecules without
colliding with the substrate, as well as sensitive enough to
measure the small changes of the force gradient ($\approx$ 0.1
N/m). The minimum detectable force gradient using our detection
scheme is estimated to be about 0.05 N/m. The RH was slowly
decreased down to a desired value during several hours by using
desiccant material placed inside a metallic enclosure box
containing the AFM. The RH was also accurately monitored within
2\% error by a digital hygrometer. Note that since our AFM did
not employ a laser for the force-gradient measurements, there was
very low electric power dissipation ($<$ 1 nW) so that the local
temperature variations near the tip were almost negligible. The
inset of Fig. 1 shows an atomic-resolution AFM image of the mica
substrate obtained at the RH of less than 5\% (scan area of 1 nm
$\times$ 1.5 nm) \cite{SCJ03}.

\begin{figure}
\scalebox{1.0}{\includegraphics{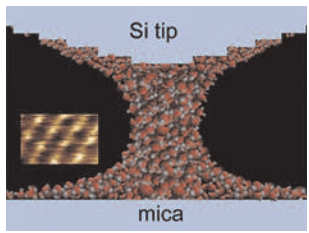}} 
\caption{Experimental schematics of a high sensitivity AFM setup
operating in ambient conditions. The inset shows an AFM image of
the dry and clean mica. }
\label{Figure1}
\end{figure}

In dynamic force microscopy, an oscillating probe detects the
force gradient between the tip and sample along its oscillation
direction. Such force gradient, $k_{ts} = - \partial F_{ts}
/\partial z$, detected by the probe is related to the resonance
frequency shift ($\Delta f$) by,~\cite{GSBL03}
\begin{equation}
k_{ts} = - \frac{\partial F_{ts}}{\partial z} = 2
k_{0}\frac{\Delta f }{f_{0}} \; . \label{eq:fshift}
\end{equation}
In our small-amplitude-modulation operation mode of AFM, however,
one has to convert the measured amplitude change to the frequency
change in order to obtain the force gradient. We have measured
such a response curve, which shows that the resonance frequency
of the tuning fork decreases by 10 Hz as the amplitude decreases
by 5\%. Thus, $\Delta f$ is related to $\Delta A \; (\equiv  A-
A_{0}$) by,
\begin{equation}
\Delta f \approx \frac{10}{0.05} \left(\frac{\Delta
A}{A_{0}}\right) \; , \label{eq:fa}
\end{equation}
where $A$ is the oscillation amplitude during the
approach/retraction process and $A_{0}$ is the free oscillation
amplitude of the tuning fork. Note that the measured frequency
shift is negative during approach and retraction, indicating that
the force gradient is also negative or the interaction force is
attractive.~\cite{Sarid94}

Figure 2 presents experimental results of the force-gradient
measurement: the magnitude of the force gradient versus the
elongation of the water column, obtained at a given RH of (a) 2\%,
(b) 15\%, (c) 31\%, and (d) 45\%. Each set of data represents one
single approach (blue dots) and retraction (red dots) process
obtained when the desired value of RH is established. The black
solid line in (b) and (c) is a guide to the eye. The origin in (a)
represents the tip-substrate contact point, whereas it denotes
the position of tip retraction in (b) to (d). The approaching tip
is computer controlled to retract immediately after its
modulation amplitude is decreased by $\approx$ 1\% due to the
effects of capillary-induced damping.

In a very dry condition of the relative humidity (RH) of 2\%, the
measured elastic force gradient versus the tip-substrate
separation showed no noticeable hysteresis behaviors during the
tip approach and retraction (Fig. 2a). The force gradient of
$\approx$ 1 N/m near the tip-substrate contact was originated from
the typical short-range ($<$ 1 nm) adhesive forces and the van
der Waals forces between the tip and the substrate. Therefore, the
absence of hysteresis in the approach-retraction curves, as well
as the atomic-resolution noncontact AFM image of the mica
substrate obtained in such a dry condition (inset of Fig. 1),
indicated that there was neither capillary condensation of water
nor any binding materials in between the tip-substrate gap. This
was expected because the mica sample was placed, after in situ
cleavage and subsequent chemical cleansing with dilute acid
\cite{KC00}, inside an air-tight enclosure \cite{footnote1}.

At 15\% RH, on the other hand, similar measurements showed quite
different results (Fig. 2b). As the tip approached the substrate,
the force gradient increased abruptly as a result of the strongly
attractive capillary forces arising from the Laplace pressure as
well as the surface tension of condensed water \cite{Is94}. Such
capillary condensation occurred at less than $\approx$ 1-nm
tip-substrate distance \cite{HXOS95,SL03}, at which there was
negligible tip-surface interaction in dry conditions (refer to
Fig. 2a). Once the nanometric condensed water meniscus was formed
and the tip modulation-amplitude is decreased by about 1\%, the
tip was computer-controlled to immediately retract in order to
keep the meniscus as thin as possible.

As can be observed in Fig. 2b, the elastic force gradient
decreased stepwise as the tip retracted in the nanometric range.
There appeared three distinct steps (although the initial
retraction region of 1.2 nm may as well be regarded as two
additional steps) until the force gradient vanished (i.e. the
water column was ruptured) at the distance of 4.2 nm from the
retraction position. In particular, there was obvious regularity
in the step heights (0.45 $\pm$ 0.03 N/m) as well as in then n
step-plateau lengths (0.6 or 1.2 nm). It was also found that the
rupture occurred abruptly within a distance of 0.1 nm, as
manifested by the steep step edge.

\begin{figure} [ht]
\scalebox{0.9}[0.8]{\includegraphics{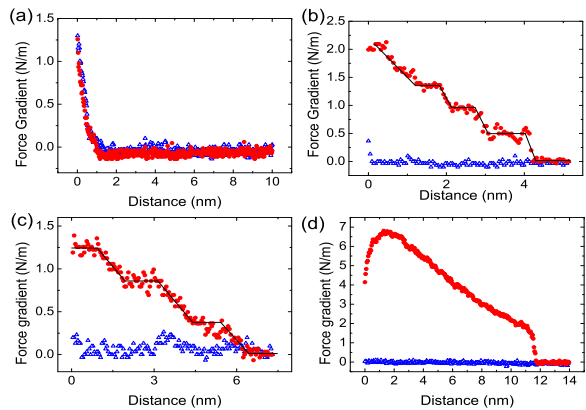}} 
\caption{Measurement of the force gradient of the water meniscus
as a function of its elongation at a given RH of (a) 2\%, (b)
15\%, (c) 31\%, and (d) 45\%. The approach and retraction speed is
0.15 nm/s.}
 \label{Figure2}
\end{figure}

The discreteness of the force gradient may be associated with the
atomic-scale stick-slip behavior \cite{DG85} of the three-phase
(i.e., tip-vapor-liquid) circular contact line developed on the
tip surface. The contact line is pinned at the atomic step
position of the discretely layered structures of the curved Si
tip surface (Fig. 1), because Si atoms at the step edge with low
bond-coordinate number exert extra attractive pinning forces on
the water molecules. Once such a pinning process occurs, the
magnitude of the force gradient in the vertical direction remains
constant during subsequent stretch of the water column, like an
elastic linear spring, as manifested by the plateaus in Fig. 2a.

While the surface area of the constant-volume water column
increases as the tip retracts \cite{footnote2}, the contact angle
with respect to the tip surface as well as the Laplace pressure
are decreased steadily \cite{Is94}. Then a slip of the contact
line occurs when the water column is stretched enough so that the
contact angle becomes equal to a critical angle, beyond which the
pinning energy can no longer compensate for the extra surface
energy needed in maintaining the same contact position. The force
gradient decreases rather abruptly during the slip process and
the slip stops when the contact angle restores its equilibrium
value at the next atomic step position of the tip where the free
energy is minimized. It is interesting that stepwise decrease
with respect to elongation was observed in the measurement of the
quantized conductance of a nanometric gold junction formed
between two gold surfaces \cite{YBBAR98}, which was shown to be
highly correlated with the stiffness change of the metallic
nanocontact \cite{BP02}.

The area under the last plateau before the rupture ($\approx$ 600
pN) represents the capillary force at the last stick position of
$\approx$ 3.1 nm. The corresponding work done until the rupture
occurs ($\approx$ 2.3 eV) is approximately equivalent to the
overall loss of tens of hydrogen bonds during the last 1-nm
elongation. A simple estimate of the meniscus-neck diameter,
given the vertical force of 600 pN, presents a value of $\approx$
2.6 nm, assuming the bulk value for the surface tension of the
water-air interface (72.8 mN/m at 20 $^{\rm o}$C)
\cite{footnote4} [note that simulation predicts this bulk value
might be reduced by $\sim$ 20\% in the nanometer dimensions
\cite{ZBL97}].

Such a thin water meniscus with $\approx$ 5 nm in length (Fig. 2b)
corresponds to a cylindrical liquid volume of $\approx$ $10^{-20}$
cm$^3$ (or 10 yoctoliter), consisting of only $\sim$ $10^3$ water
molecules. As the meniscus thins even more, the relative
fluctuation of the liquid-vapor interface increases so that the
meniscus becomes thermodynamically unstable \cite{WPLT00}. The
observed rupture of the meniscus occurs when the neck has a width
of $\le$ 2.6 nm and this critical value is in agreement with the
simulation results \cite{JSR02}. Such interfacial instability
also manifests itself as an increased level of fluctuations of
the force gradient in the last plateau before the rupture (Fig.
2b).

The results at 31\% RH (Fig. 2c), which were independently
obtained with the same tip on a different day, showed overall
step-like discrete characteristics similar to those in Fig. 2b,
except that the step edges following the plateaus were broader
(0.9 $\pm$ 0.1 nm). At a higher RH of 45\% (Fig. 2d), the
elasticity-distance curves exhibited smooth variations except for
the final rupture process, as similarly observed in Ref.
\cite{ZHM02}. The continuous and monotonic decrease of the
elastic-force gradient as well as the much larger initial values
of the force gradient indicated that the water meniscus was
already of bulk-like nature before retraction, smearing out the
stepwise discrete changes \cite{BR02}.

\begin{figure} [t]
\begin{center}
\scalebox{0.8}[0.87]{\includegraphics{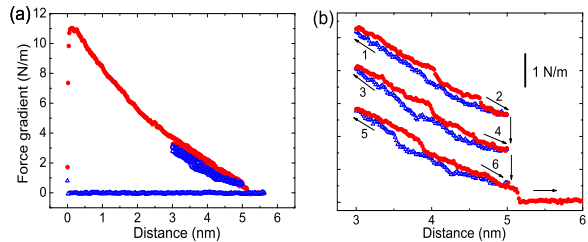}} 
\caption{Reproducibility of the repeated short-ranged
approach-retraction cycles of the force-gradient measurement,
obtained at 13\% RH. (a) The tip makes a deeper initial approach
followed by repetitive cycles of retraction and approach until the
water bridge is intentionally ruptured. (b) Close-up plot of the
repeated cycles, with an arbitrary vertical offset for an eye
guide. The approach and retraction speed of the tip is 0.15 nm/s.
\label{Figure3}}
\end{center}
\end{figure}

In other words, as the initial amount of capillary-condensed
water increased due to a uniform water layer formed on the mica
substrate at an RH higher than 40\% \cite{HXOS95}, the contact
circle became larger in diameter and the effective pinning
strength was diminished, resulting in the usual steady sliding of
bulk water. The broadened step edges shown in Fig. 2c may be
caused by mixed effects of the stick-slip and the steady sliding
behaviors. Note that the elasticity measurements in Fig. 2 were
performed during a single approach-retraction process, after
waiting for many hours until the RH reached a given value
\cite{footnote3}, in order to avoid any unwanted `memory' effects
of the water meniscus as well as to form the thinnest possible
meniscus width at a given RH.

To confirm that the measured force gradient during approach and
retraction of the tip truly represented the elasticity of the
system, and to explore the potential of the present technique for
manipulation of the nanometric aqueous column, we repeatedly
stretched and relaxed the water column by a slow (0.15 nm/s
speed) cyclic motion of the tip. Figure 3a shows an
approach-retraction curve, which was obtained when the approaching
tip retracts as soon as its initial modulation amplitude is
decreased by $\approx$ 10\%, resulting in a `closer' approach with
respect to that in Fig. 2b obtained at a similar RH (note that the
tip approached only until its modulation amplitude is decreased
by $\approx$ 1\% in Fig. 2b). The repeated cycles then start when
the tip retracts from the origin up to $\approx$ 5-nm distance
just before the rupture, and continue within the subsequent 2-nm
range.

Figure 3b presents a close-up plot of the repeated cycles in Fig.
3a, with an arbitrary vertical offset given between each cycle
just for a clear eye guide. The vertical scale bar represents the
force gradient of 1 N/m. The tip movement is in the order of the
increasing arrow number. The retraction curve of the last cycle
(arrow $\#$6) is followed by an intentional rupture of the thin
water column. The evident reproducibility of the repeated cycles
in Fig. 3b represents that the force-gradient measurements indeed
reveal the discrete elastic properties of the thin water column
confined in between the tip-substrate gap. In particular, the
repeated hysteresis behaviors within each cycle are indicative of
the atomic-scale contact angle hysteresis of the nanometric liquid
water on the tip surface \cite{DG85}, which is closely related to
the atomic-scale friction.

The present work may provide a novel experimental tool for
studying the kinematics of the condensed or adsorbed liquids on
surfaces, which is of fundamental and technological interest in
surface science and engineering \cite{PZXHM99}. One can also
study water molecules in nanoconfinement \cite{PS02} that may
behave unexpectedly by forming nanoclusters or by rearranging
themselves to seek the energetically most favorable
configurations by bending, but not breaking, the finite
hydrogen-bonded network \cite{BD94}. Furthermore, the current
technique to form and manipulate a nanometric water column
between two surfaces may provide a means to extend our
understanding of the transport processes of ions through a
nanometric water channel in biological cells.


\acknowledgements We are grateful to H. E. Stanley, S. Buldyrev,
J. Yu, Y.-W. Son, K. Kim, M. H. Lee, and J. Jang for helpful
discussions. This work was supported by the Korean Ministry of
Science and Technology through Creative Research Initiatives
Program.


\bibliography{referencewater}

\begin{thebibliography}{29}
\bibitem[{foo({\natexlab{a}})}]{footnote1}
\bibinfo{note}{The experimental results of the force-gradient measurements
  presented in this work did not show any noticeable dependence on whether the
  substrate was chemically treated or not, which indicated that the measured
  elasticity properties associated with the thin water bridge were not
  sensitive to any possible surface contaminants or impurities.}
\bibitem[{foo({\natexlab{b}})}]{footnote2}
\bibinfo{note}{The initial width of the water meniscus depended much upon the
  rapidity of the tip retraction after the abrupt development of capillary
  condensation at a given RH. Once the initial meniscus was formed thereby, its
  volume remained preserved during subsequent tip retraction and the overall
  decrease of the force gradient with the elongation of the water column may be
  attributed to the consequent reduction of the meniscus width.}
\bibitem[{foo({\natexlab{c}})}]{footnote4}
\bibinfo{note}{The additional contribution of the Laplace pressure to the force
  was negligible before the rupture, which resulted in only a slight reduction
  of the estimated neck diameter.}
\bibitem[{foo({\natexlab{d}})}]{footnote3}
\bibinfo{note}{Similar force-gradient measurements, performed after more rapid
  (at 1 nm/s) and repeated (30 times) cycles were executed, showed very
  different behaviors of a longer plateau ($\sim$ 10 nm) both in the approach
  and the retraction curve, before the rupture at a larger force-gradient
  ($\sim 10$ N/m) occurred. This observation may be associated with the fact
  that a bulk-like meniscus with a submicrometre-scale thickness \cite{SFR98}
  was produced due to the tip being dipped repeatedly in a microscopic water
  droplet ($\sim$ 10 nm height and 1 $\mu$m base-diameter) formed on the mica
  substrate \cite{XLHOS98}.}
\end{thebibliography}

\end{document}